\documentclass[final,nofootinbib,aps,twocolumn,showkeys,showpacs,groupaddress,preprintnumbers,floatfix]{revtex4-1}

\usepackage{dynlearn}

\newcommand{\anddist}{
  \begin{tabular}[t]{cccc}
    \multicolumn{4}{c}{\textsc{And}}                            \\
    \toprule
    $\source{0}$ & $\source{1}$ & $\target$ & $\Pr$             \\
    \midrule
    0            & 0            & 0         & $\nicefrac{1}{4}$ \\
    0            & 1            & 0         & $\nicefrac{1}{4}$ \\
    1            & 0            & 0         & $\nicefrac{1}{4}$ \\
    1            & 1            & 1         & $\nicefrac{1}{4}$ \\
    \bottomrule
  \end{tabular}
}

\newcommand{\propname}[1]{\hfill \phantom{a} \hfill \emph{(#1)}}
\newcommand{\powerset} [1] {\ensuremath{\mathcal{P}^{+}\!\!\left(#1\right)}\xspace}

\newcommand{\sep}{\cdot}

\newcommand{\source} [1] {\ensuremath{X_{#1}}\xspace}
\newcommand{\target} {\ensuremath{Y}\xspace}

\newcommand{\PID} [3] {
  \ensuremath{
    \operatorname{I_{#1}}
    \if\relax\detokenize{#3}\relax
      \!\!
    \else
      \left[ #2 \rightarrow #3 \right]
    \fi
  }
  \xspace
}

\newcommand{\Icap}  [2] [\target] {\PID{\cap}{#2}{#1}}
\newcommand{\Ipart} [2] [\target] {\PID{\partial}{#2}{#1}}
\newcommand{\Imin}  [2] [\target] {\PID{\textrm{min}}{#2}{#1}}
\newcommand{\Iproj} [2] [\target] {\PID{\textrm{proj}}{#2}{#1}}
\newcommand{\Ibroja}[2] [\target] {\PID{\textrm{BROJA}}{#2}{#1}}
\newcommand{\Iccs}  [2] [\target] {\PID{\textrm{ccs}}{#2}{#1}}
\newcommand{\Idep}  [2] [\target] {\PID{\textrm{dep}}{#2}{#1}}
\newcommand{\Immi}  [2] [\target] {\PID{\textrm{mmi}}{#2}{#1}}
\newcommand{\Ida}   [2] [\target] {\PID{\downarrow}{#2}{#1}}
\newcommand{\Iwedge}[2] [\target] {\PID{\wedge}{#2}{#1}}

\colorlet{xycolor}  {blue}
\colorlet{xzcolor}  {green!66!black}
\colorlet{yzcolor}  {red}
\colorlet{xyzcolor} {ryan}

\begin{document}

\def\ourTitle{
  Unique Information via Dependency Constraints
}

\def\ourAbstract{
The partial information decomposition (PID) is perhaps the leading proposal for
resolving information shared between a set of sources and a target into
redundant, synergistic, and unique constituents. Unfortunately, the PID
framework has been hindered by a lack of a generally agreed-upon, multivariate
method of quantifying the constituents. Here, we take a step toward rectifying
this by developing a decomposition based on a new method that quantifies unique
information. We first develop a broadly applicable method---the dependency
decomposition---that delineates how statistical dependencies influence the
structure of a joint distribution. The dependency decomposition then allows us
to define a measure of the information about a target that can be uniquely
attributed to a particular source as the least amount which the source-target
statistical dependency can influence the information shared between the sources
and the target. The result is the first measure that satisfies the core axioms
of the PID framework while not satisfying the Blackwell relation, which depends
on a particular interpretation of how the variables are related. This makes a
key step forward to a practical PID.
}

\def\ourKeywords{
  partial information decomposition, mutual information, statistical dependence, information theory, cybernetics.
}

\hypersetup{
  pdfauthor={James P. Crutchfield},
  pdftitle={\ourTitle},
  pdfsubject={\ourAbstract},
  pdfkeywords={\ourKeywords},
  pdfproducer={},
  pdfcreator={}
}

\author{Ryan G. James}
\email{rgjames@ucdavis.edu}

\author{Jeffrey Emenheiser}
\email{jemenheiser@ucdavis.edu}

\author{James P. Crutchfield}
\email{chaos@ucdavis.edu}
\affiliation{Complexity Sciences Center and Physics Department, University of California at Davis, One Shields Avenue, Davis, CA 95616}

\date{\today}
\bibliographystyle{unsrt}

\title{\ourTitle}

\begin{abstract}
\ourAbstract
\end{abstract}



\preprint{\sfiwp{17-09-034}}
\preprint{\arxiv{1709.06653}}

\title{\ourTitle}

\date{\today}

\maketitle

\setstretch{1.1}

\section{Introduction}
\label{sec:introduction}

Understanding how information is stored, modified, and transmitted among the components of a complex system is fundamental to the sciences.
Application domains where this would be particularly enlightening include gene regulatory networks~\cite{gates2016control}, neural coding~\cite{faber2018computation}, highly-correlated electron systems, spin lattices~\cite{vijayaraghavan2017anatomy}, financial markets~\cite{james2018modes}, network design~\cite{arellano2013shannon}, and other complex systems whose large-scale organization is either not known a priori or emerges spontaneously.
Information theory's originator Claude Shannon~\cite{Shan48a} was open to the possible benefits of such applications; he was also wary \cite{Shan56b}.
In an early attempt to lay common foundations for multicomponent, multivariate information Shannon~\cite{Shan53a} appealed to Garrett Birkhoff's lattice theory~\cite{Birk40a}.
Many of the questions raised are still open today~\cite{Will10a,james2017multivariate}.

Along these lines, but rather more recent, one particularly promising framework
for accomplishing such a decomposition is the \emph{partial information
decomposition} (PID)~\cite{Will10a}.
Once a practitioner partitions a given set of random variables into \emph{sources} and a \emph{target}, the framework decomposes the information shared between the two sets into interpretable, nonnegative components---in the case of two sources: redundant, unique, and synergistic informations.
This task relies on two separate aspects of the framework: first, the overlapping source subsets into which the information should be decomposed and, second, the method of quantifying those informational components.

Unfortunately, despite a great deal of effort~\cite{harder2013bivariate,griffith2014quantifying,bertschinger2014quantifying,chicharro2017quantifying,griffith2014intersection,ince2017measuring,bertschinger2013shared,rauh2014reconsidering,chicharro2016redundancy,banerjee2015synergy}, the current consensus is (i) that the lattice needs to be modified~\cite{rauh2017secret,chicharro2016redundancy,rauh2017extractable,bertschinger2013shared,rauh2014reconsidering} and (ii) that extant methods of quantifying informational components~\cite{Will10a,harder2013bivariate,bertschinger2014quantifying,griffith2014quantifying,griffith2014intersection,ince2017measuring} are not satisfactory in full multivariate generality due to either only quantifying unique informations, being applicable only to two-source distributions, or lacking nonnegativity.
Thus, the promise of a full informational analysis of the organization of complex systems remains unrealized after more than a half century.

The following addresses the second aspect---quantifying the components.
Inspired by early cybernetics---specifically, Krippendorff's lattice of system models (reviewed in Ref.~\cite{krippendorff2009ross})---we develop a general technique for decomposing arbitrary multivariate information measures according to how they are influenced by statistical dependencies.
\footnote{Since the development of this manuscript, it has come to the authors' attention that this structure had been independently developed within the field of system science~\cite{zwick2004overview}.}
We then use this decomposition to quantify the information that one variable uniquely has about another.
Reference~\cite{bertschinger2014quantifying}'s \Ibroja[]{} measure also
directly quantifies unique information. However, depending upon one's
intuitions~\cite{james2018perspective} it can be seen to inflate redundancy~\cite{ince2017measuring}.
Both our measure as well as Ref.~\cite{ince2017measuring}'s \Iccs[]{} take
into account the joint statistics of the sources, but \Iccs[]{} does so at the expense of positivity.
This makes our proposal the only method of quantifying the partial information decomposition that is nonnegative, respects the source statistics, and satisfies the core axioms of the PID framework.

Our development proceeds as follows.
Section\nobreakspace \ref {sec:background} reviews the PID and Section\nobreakspace \ref {sec:measuring} introduces our measure of unique information.
Section\nobreakspace \ref {sec:examples} then compares our measure to others on a variety of exemplar distributions, exploring and contrasting its behavior.
Section\nobreakspace \ref {sec:discussion} discusses several open conceptual issues and Section\nobreakspace \ref {sec:conclusion} concludes.
The development requires a working knowledge of information theory, such as found in standard texts~\cite{Cover2006,MacKay2003,Yeung2008}.

\section{Background}
\label{sec:background}

Consider a set of \emph{sources} $\source{0}, \source{1}, \ldots, \source{n-1} = \source{0:n}$ and a \emph{target} \target.\footnote{We subscript the joint variable with a Python-like array-slice notation, which matches Dijkstra's argument~\cite{dijkstra1982numbering}.}
The amount of information the sources carry about the target is quantified by their mutual information:
\begin{align*}
  \I{\source{0:n} : \target} &= \I{\source{0}, \source{1}, \ldots, \source{n-1} : \target} \\
                             &= \sum p(\source{0:n}, \target) \log_2 \frac{p(\source{0:n}, \target)}{p(\source{0:n})p(\target)}
  ~.
\end{align*}
The PID then assigns \emph{shared information} to sets of source groupings such that no (inner) set is subsumed by another~\cite{Will10a}.
In this way, the PID quantifies what information about the target each of those groups has in common.

\subsection{Antichain Lattices}
\label{sec:Antichain}

\begin{figure}
  \centering
  \includegraphics{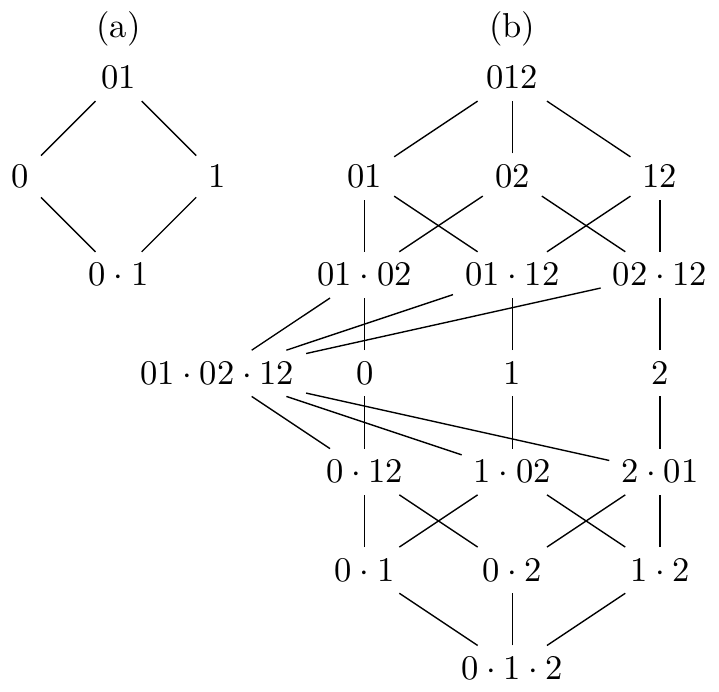}
\caption{
  Lattice of antichains for (a) two ($\source{0}$ and $\source{1}$) and (b) three sources ($\source{0}$, $\source{1}$, and $\source{2}$): An antichain is represented using a dot to separate sets and sets by concatenated indices; \eg, $\left\{\left\{\source{0}\right\}\left\{\source{1},\source{2}\right\}\right\}$ is represented $0\sep12$.
	}
\label{fig:lattices}
\end{figure}

The sets of groupings we consider are \emph{antichains}:
\begin{align*}
  \mathcal{A}(\source{0:n}) = \left\{ \alpha \in \powerset{\powerset{\source{0:n}}}
    : \forall s_1, s_2 \in \alpha, s_1 \not\subset s_2 \right\}
  ,
\end{align*}
where $\powerset{S} = \mathcal{P}(S) \setminus \{\emptyset\}$ denotes the set of nonempty subsets of set $S$.
Antichains form a lattice \cite{Birk40a}, where one antichain $\alpha$ is less than another $\beta$ if each element in $\beta$ subsumes some element of $\alpha$:
\begin{align*}
  \alpha \preceq \beta \iff \forall s_1 \in \beta, \exists s_2 \in \alpha, s_2 \subseteq s_1
  ~.
\end{align*}
Figure\nobreakspace \ref {fig:lattices} graphically depicts antichain lattices for two and three variables.
There, for brevity's sake, a dot separates the sets within an antichain, and the groups of sources are represented by their indices concatenated.
For example, $0\sep12$ represents the antichain $\left\{\left\{\source{0}\right\}\left\{\source{1}, \source{2}\right\}\right\}$.

\subsection{Shared Informations}
\label{sec:SharedInfo}

Given the antichain lattice, one then assigns a quantity of shared or redundant information to each antichain.
This should quantify the amount of information shared by each set of sources within an antichain $\alpha$ about the target.
This shared information will be denoted \Icap{\alpha}\!\!\!.
Reference~\cite{Will10a} put forth several axioms that such a measure should follow:
\begin{itemize}
\item[\textbf{(S)}] \label{axiom:s}
  $\Icap{\alpha}$ is unchanged under permutations of $\alpha$.
	\propname{symmetry}
\item[\textbf{(SR)}] \label{axiom:sr}
	$\Icap{i} = \I{\source{i} : \target}$.
	\propname{self-redundancy}
\item[\textbf{(M)}] \label{axiom:m}
	For all $\alpha \preceq \beta$, $\Icap{\alpha} \leq \Icap{\beta}$.
	\propname{monotonicity}
\end{itemize}

With a lattice of shared informations in hand, the \emph{partial information} \Ipart{\alpha} is defined as the M{\"o}bius inversion~\cite{Birk40a} of the shared information:
\begin{align}
  \Icap{\alpha} = \sum_{\beta \preceq \alpha} \Ipart{\beta}
  ~.
  \label{eq:mobius}
\end{align}
We further require that the following axiom hold:
\begin{itemize}
\item[\textbf{(LP)}] \label{axiom:lp} $\Ipart{\alpha} \geq 0$.
	\propname{local positivity}
\end{itemize}
This ensures that the partial information decomposition forms a partition of the sources-target mutual information and contributes to the decomposition's interpretability.

\subsection{The Bivariate Case}
\label{subsec:bivariate}

In the case of two inputs, the PID takes a particularly intuitive form.
First, following the self-redundancy axiom \textbf{(SR)}, the sources-target mutual information decomposes into four components:
\begin{align}
  \I{\source{0} \source{1} : \target} &= \Ipart{0\sep1}
                                       + \Ipart{0} \nonumber\\
                                      &\quad + \Ipart{1}
                                             + \Ipart{01}
  \label{eq:bivariatea}
  ~,
\end{align}
and, again following \textbf{(SR)}, each source-target mutual information consists of two components:
\begin{align}
  \I{\source{0} : \target} &= \Ipart{0\sep1} + \Ipart{0} \label{eq:bivariateb} \\
  \I{\source{1} : \target} &= \Ipart{0\sep1} + \Ipart{1} \label{eq:bivariatec}
  ~.
\end{align}

The components have quite natural interpretations.
\Ipart{0\sep1} is the amount of information that the two sources \source{0} and \source{1} \emph{redundantly} carry about the target \target.
\Ipart{0} and \Ipart{1} quantify the amount of information that sources \source{0} and \source{1}, respectively, carry \emph{uniquely} about the target \target.
Finally, \Ipart{01} is the amount of information that sources \source{0} and \source{1} \emph{synergistically} or collectively carry about the target \target.

Combining the above decompositions, we see that the operational result of conditioning removes redundancy but expresses synergistic effects:
\begin{align*}
  \I{\source{0} : \target | \source{1}} &= \I{\source{0} \source{1} : \target}
                                           - \I{\source{1} : \target} \\
                    &= \Ipart{0} + \Ipart{01}
  ~.
\end{align*}
Furthermore, the co-information~\cite{Bell2003} can be expressed as:
\begin{align*}
  \I{\source{0} : \source{1} : \target} &= \I{\source{0} : \target} - \I{\source{0} : \target | \source{1}} \\
  &= \Ipart{0 \sep 1} - \Ipart{01}
  ~.
\end{align*}
This illustrates one of the PID's strengths.
It explains, in a natural fashion, why the co-information can be negative.
It is the difference between a distribution's redundancy and synergy.

The bivariate decomposition's four terms are constrained by the three
self-redundancy constraints Eqs.\nobreakspace  \textup {(\ref {eq:bivariatea})} to\nobreakspace  \textup {(\ref {eq:bivariatec})} .
This leaves one degree of freedom. Generally, though not always~\cite{bertschinger2014quantifying}, this is taken as \Ipart{0\sep1}.
Therefore, specifying any component of the partial information lattice determines the entire decomposition.
In the multivariate case, however, no single \Ipart{\alpha} (redundancy, unique, synergistic, or otherwise), when combined with relations to standard information-theoretic quantities, will determine the remainder of the values.
For this reason, one generally relies upon quantifying the \Icap[]{} values to complete the decomposition via the M{\"o}bius inversion.

Finally, in the bivariate case one further axiom has been
suggested~\cite{harder2013bivariate}, though not put forth in original PID:
\begin{itemize}
\item[\textbf{(Id)}] \label{axiom:id}
	$\Icap[\source{0}\source{1}]{0 \sep 1} = \I{\source{0} : \source{1}}$
	\propname{identity}
\end{itemize}
This axiom ensures that simply concatenating independent inputs does not result in redundant information.
The identity axiom, though intuitive in the case of two inputs, suffers from several issues.
Primarily, with three or more sources it is known to be inconsistent with local positivity \textbf{(LP)}.
Furthermore, it is not clear how to extend this axiom to the multivariate case or even if it should be extended.
In short, though many proposed methods of quantifying the PID satisfy the identity axiom, it is certainly not universally accepted.

\subsection{Extant Methods}
\label{subsec:extant}

Several methods can be easily set
aside---\Immi[]{}~\cite{bertschinger2013shared},
\Iwedge[]{}~\cite{griffith2014intersection}, and
\Ida[]{}~\cite{griffith2014quantifying,bertschinger2013shared}---as suffering from significant drawbacks.
\Immi[]{} necessarily assigns a zero value to at least one of the unique
informations, doing so by dictating that whichever source shares the least
amount of information with the target, it does so entirely redundantly with the other sources.
\Iwedge[]{} is based on the G{\'a}cs-K{\"o}rner common information. And, so it
is insensitive to any sort of statistical correlation that is not a common random variable.
\Ida[]{} quantifies unique information from each source directly using an upper
bound on the secret key agreement rate~\cite{maurer1999unconditionally}, but in
a way that leads to inconsistent redundancy and synergy values.

Now, we can turn to describe the four primary existing methods for quantifying the PID.
\Imin[]{}, the first measure proposed~\cite{Will10a}, quantifies the average least information the individual sources have about each target value.
It has been criticized~\cite{griffith2014quantifying,harder2013bivariate} for its behavior in certain situations.
For example, when the target simply concatenates two independent bit-sources, it decomposes those two bits into one bit of redundancy and one of synergy.
This is in stark contrast to the more intuitive view that the target contains two bits of unique information---one from each source.

\Iproj[]{} quantifies shared information using information geometry \cite{harder2013bivariate}.
Due to its foundation relying on the Kullback-Leibler divergence, however, it does not naturally generalize to measuring the shared information in antichains of size three or greater.

\Ibroja[]{} attempts to quantify unique
information~\cite{bertschinger2014quantifying}, as does our approach.
It does this by finding the minimum \I{\source{i} : \target | \source{0:n \setminus i}} over all distributions that preserve source-target marginal distributions.
(The random variable set \source{0:n \setminus i}, excludes variable $\source{i}$.)
Depending on one's perspective on the roles of source and target variables, however, \Ibroja[]{} can be seen to artificially correlate the sources and thereby overestimate redundancy~\cite{ince2017measuring,james2018perspective}.
This leads the measure to quantify identical, though independent, source-target channels as fully redundant.
Furthermore, as a measure of unique information, it cannot completely quantify the partial information lattice when the number of sources exceeds two.

Finally, \Iccs[]{} quantifies redundant information by aggregating the pointwise coinformation terms whose signs agree with the signs of all the source-target marginal pointwise mutual informations \cite{ince2017measuring}.
This measure seems to avoid the issue used to criticize \Imin[]{}, that of
capturing the ``same quantity'' rather than the ``same information''.
Further, it respects the source statistics unlike \Ibroja[]{}.
Notably, it can be applied to antichains of any size.
Unfortunately, it does so incurring the expense of negativity, though one can argue that this is an accurate assessment of information architecture.

With these measures, their approaches, and their limitations in mind, we now turn to defining our measure of unique information.

\section{Unique Information}
\label{sec:measuring}

We now propose a method to quantify partial information based on terms of the
form \Ipart{i}\!\!\!\!---that is, the unique information. We begin by discussing the notion of \emph{dependencies} and how to quantify their influence on information measures.
We then adapt this to quantify how source-target dependencies influence the sources-target mutual information.
Our measure defines unique information \Ipart{i} as the least amount that the
$\source{i} \target$ dependency can influence the shared information \I{\source{0:n} : \target}.

\subsection{Constraint Lattice}
\label{subsec:lattice}

We begin by defining the \emph{constraint lattice} $\mathcal{L}(\Sigma)$, a lattice of sets of subsets of variables.
These subsets of variables are antichains, as in the partial information
lattice, but are further constrained and endowed with a different ordering.
Specifically, given a set of variables $\Sigma = \left\{\source{0}, \source{1}, \ldots\right\}$, a \emph{constraint} $\gamma$ is a nonempty subset of $\Sigma$.
And, a \emph{constraint set} $\sigma$ is a set of constraints that form an antichain on $\Sigma$ and whose union covers $\Sigma$; they are \emph{antichain covers}~\cite{Birk40a}.
Concretely, $\sigma \in \mathcal{P}^{+}(\mathcal{P}^{+}(\Sigma))$ such that, for all $\gamma_1, \gamma_2 \in \sigma$, $\gamma_1 \nsubseteq \gamma_2$ and $\bigcup \sigma = \Sigma$.
The constraint sets are required to be covers since we are not concerned with each individual variable's distribution, rather we are concerned with how the variables are related.
We refer to these variable sets as constraints since we work with families of distributions for which marginal distributions over the variable sets are held fixed.

\begin{figure}
  \includegraphics{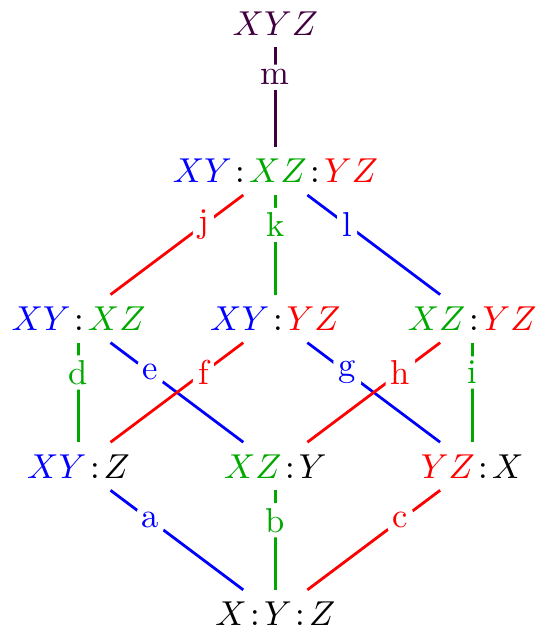}
\caption{
  Constraint lattice of three random variables $X$, $Y$, and $Z$.
  Blue edges (a, e, g, l) correspond to adding constraint $XY$, green (b, d, i, k) to adding $XZ$, and red (c, f, h, j) to adding $YZ$.
	}
\label{fig:constraintlattice}
\end{figure}

There is a natural partial order $\sigma_1 \preceq \sigma_2$ over constraint sets:
\begin{align*}
  \sigma_1 \preceq \sigma_2 \iff \forall \gamma_1 \in \sigma_1, \exists \gamma_2 \in \sigma_2, \gamma_1 \subseteq \gamma_2
  ~.
\end{align*}
Note that this relation is somewhat dual to that in the PID and, furthermore, that the set of antichain covers is a subset of all antichains.
The lattice $\mathcal{L}(\Sigma)$ induced by the partial order on $\Sigma = \left\{X, Y, Z\right\}$ is displayed in Fig.\nobreakspace \ref {fig:constraintlattice}.
The intuition going forward is that each node (antichain) in the lattice represents a set of constraints on marginal distributions and the constraints at one level imply those lower in the lattice.

\subsection{Quantifying Dependencies}
\label{subsec:dependencies}

To quantify how each constraint set influences a distribution $p$, we associate a maximum entropy distribution with each constraint set $\sigma$ in the lattice.
Specifically, consider the set $\Delta_p(\sigma)$ of distributions that match marginals in $\sigma$ with $p$:
\begin{align}
\Delta_p(\sigma)
    = \left\{ q : p(\gamma) = q(\gamma), ~\gamma \in \sigma \right\}
  ~.
\label{eq:distconstraints}
\end{align}
To each constraint set $\sigma$ we associate the distribution in $\Delta_p(\sigma)$ with maximal Shannon entropy:
\begin{align}
  p_\sigma = \arg\max \left\{ \H{q} : q \in \Delta_p(\sigma) \right\}
  ~.
  \label{eq:argmaxent}
\end{align}
This distribution includes no additional statistical structure beyond that constrained by $\sigma$~\cite{Jayn83}.

When an information measure, such as the mutual information, is computed
relative to the maximum entropy distribution $p_\sigma$, we subscript it with
the constraint $\sigma$.
For example, the mutual information between the joint variable $XY$ and the
variable $Z$ relative the the maximum entropy distribution satisfying the
constraint $XY:YZ$ is denoted $\I[XY:YZ]{X Y : Z}$.

Given this lattice of maximum entropy distributions, we can then compute any multivariate information measure on those distributions and analyze how its value changes moving across the lattice.
Moves here correspond to adding or subtracting dependencies.
We call the lattice of information measures applied to the maximum entropy distributions the \emph{dependency structure} of distribution $p$.

The dependency structure of a distribution is a broadly applicable and robust method for analyzing how the structure of a distribution affects its information content.
It is effectively a partial order on a multiverse associated with $p$:
Consider every possible alternative universe in which select statistical
dependencies are removed from $p$.
It allows each dependency to be studied in the context of other dependencies,
leading to a vastly more nuanced view of the interactions among $p$'s variables.
We believe it will form the basis for a wide variety of information-theoretic dependency analyses in the future.

We note that this dependency structure was independently announced~\cite{virgodecomposing} after a preprint (arxiv.org:1609.01233) of the present work appeared.
There, dependency structure minimizes the Kullback-Leibler divergence, which is known to be equivalent to our maximum entropy approach~\cite{amari2001information}.
A decomposition of total correlation $\sum_j \H{X_i} - \H{X_1,\ldots,X_n}$ was studied that, in the case of three variables, amounts to decomposing each conditional mutual information into two components: e.g., $\I{X:Y|Z} = \DKL{XY:XZ:YZ}{XZ:YZ} + \DKL{XYZ}{XY:XZ:YZ}$, where the latter is the third-order connected information~\cite{Schneidman2003} and $\DKL{P}{Q}$ is the Kullback-Liebler divergence~\cite{Cover2006}.
In contrast to a lattice decomposition of total correlations, the primary contribution here applies any desired information measure to each node in the dependency structure.
This leads to a vast array of possible analyses.

As an example, consider the problem of determining ``causal pathways'' in a
network~\cite{runge2015quantifying,sun2014causation}.\footnote{The term
``causal'' here is unfortunate, due to the great deal of debate on
determining causality without intervention~\cite{Pear88a}.}
Given two paths between two network nodes, say $A \to B \to C$ and $A \to B^\prime \to C$, one would like to determine through which pathway $A$'s behavior most strongly influences $C$.
This pathway is termed the causal pathway.
Na{\"i}vely, one might assume that the pathway whose links are strongest is the more influential pathway or that it is the pathway maximizing a multivariate information measure.
Consider, however, the case where $A$ strongly influences one aspect of $B$ while another, independent aspect of $B$ strongly influences $C$.
Here, $A$ would have no influence whatsoever upon $C$ in spite of the strong individual links.
Our dependency structure, in contrast, would easily detect this via $\I[AB:BC]{A : C} = 0$, which quantifies exactly how much $A$ and $C$ are correlated through the pathway $A \to B \to C$.
In this fashion, determining causal pathways in networks becomes straightforward: For each potential pathway, consider the constraint consisting of all its links and evaluate the influence measure of choice (time-delayed mutual information, transfer entropy, or similar) between the beginning and end of the pathway.
The value indicates the pathway's strength as quantified by the chosen measure (and whose interpretation is dependent on the chosen measure).
While this example does not use the full dependency structure, it does demonstrate the usefulness of considering information measures in the context of only specific dependencies.
Furthermore, there exist other information-based methods of determining causal pathways~\cite{runge2015quantifying}, however this provides a novel and independent method of doing so.

Another application is the determination of drug interactions.
Given a dataset of responses to a variety of drugs, one would like to determine which subsets of drugs interact with one another.
One method of doing so would be to construct the dependency structure, quantifying each node with the entropy.
Then, the lowest node in the lattice whose entropy is ``close enough'' (as
determined by context) to that of the true distribution contains the minimal
set of constraints that give rise to the full structure of the true distribution.
That minimal set of constraints determines the subset of variables that are necessarily interacting.
Note that this is an application of reconstructability
analysis~\cite{zwick2004overview} and does not use the flexibility of employing a variety of information measures on the lattice.

\subsection{Quantifying Unique Information}
\label{subsec:unique}

To measure the unique information that a source---say, \source{0}---has about the target \target, we use the dependency decomposition constructed from the mutual information between sources and the target.
Consider further the lattice edges that correspond to the addition of a particular constraint:
\begin{align*}
E(\gamma) = \left\{
  \left(\sigma_1, \sigma_2\right) \in \mathcal{L}:
  \gamma \in \sigma_1,
  \gamma \notin \sigma_2 \right\}
  ~.
\end{align*}
For example, in Fig.\nobreakspace \ref {fig:constraintlattice}'s constraint lattice $E(XY)$ consists of the following edges: $(XY\!:\!Z, X\!:\!Y\!:\!Z)$, $(XY\!:\!XZ, XZ\!:\!Y)$, $(XY\!:\!YZ, YZ\!:\!X)$, and $(XY\!:\!XZ\!:\!YZ, XZ\!:\!YZ)$.
These edges---labeled $a$, $d$, $g$, and $l$---are colored blue there.
We denote a change in information measure along edge $(\sigma_1, \sigma_2)$ by $\Delta^{\sigma_1}_{\sigma_2}$.
For example, $\Delta^{\sigma_1}_{\sigma_2} \I{X Y : Z} = \I[\sigma_1]{X Y : Z} - \I[\sigma_2]{X Y : Z}$.

Our measure $\Idep{i}$ of \emph{unique information} from variable $X_i$ to the target $\target$ is then defined using the lattice $\mathcal{L}(\source{i}, \target, \source{0:n \setminus i})$:
\begin{align*}
\Idep{i} &= \min_{(\sigma_1, \sigma_2) \in E(X_i Y)}
  \left\{ \Delta^{\sigma_1}_{\sigma_2} \I{\source{0:n} : \target} \right\} \\
  &= \min
  \begin{cases}
    \I[\source{i}\target:\source{0:n \setminus i}\target]{\source{i} : \target | \source{0:n \setminus i}} & \\
    \I[\source{i}\source{0:n \setminus i}:\source{i}\target:\source{0:n \setminus i}\target]{\source{i} : \target | \source{0:n \setminus i}} &
  \end{cases}
  ~\!\!\!\!\!\!\!\!.
\end{align*}
That is, the information learned uniquely from \source{i} is the least change in sources-target mutual information among all the edges that involve the addition of the $\source{i} \target$ constraint.
Due to information-theoretic constraints (see Appendix\nobreakspace \ref {app:constrained}), the edge difference achieving the minimum must be one of either $\I[\source{i}\target:\source{0:n \setminus i}\target]{\source{i} : \target | \source{0:n \setminus i}}$ or $\I[\source{i}\source{0:n \setminus i}:\source{i}\target:\source{0:n \setminus i}\target]{\source{i} : \target | \source{0:n \setminus i}}$.
This latter quantity arises in directed reconstructability analysis~\cite{zwick2004overview}, where it has an interpretation similar to unique information.
It would not, however, result in a PID that satisfied \textbf{(LP)}; though,
when combined as above with the first quantity, local positivity is preserved.
In the case of bivariate inputs, this measure of unique information results in a decomposition that satisfies \textbf{(S)}, \textbf{(SR)} (by construction), \textbf{(M)}, \textbf{(LP)}, and \textbf{(Id)}, as shown in Appendix\nobreakspace \ref {app:properties}.
In the case of multivariate inputs, satisfying \textbf{(Id)} implies that \textbf{(LP)} is not satisfied. Further, it is not clear whether \Idep[]{} satisfies \textbf{(M)}.

With a measure of unique information in hand, we now need only describe how to determine the partial information lattice.
In the bivariate sources case, this is straightforward: self-redundancy \textbf{(SR)}, the unique partial information values, and the M{\"o}bius inversion Eq.\nobreakspace \textup {(\ref {eq:mobius})} complete the lattice.
In the multivariate case, completion is not generally possible. That said, in
many relatively simple cases combining monotonicity \textbf{(M)},
self-redundancy \textbf{(SR)}, the unique values, and a few heuristics allow
the lattice to be determined. Though, due to \Idep[]{} satisfying the identity axiom such values may violate local positivity.
The heuristics include using the M{\"o}bius inversion on a subset of the lattice as a linear constraint.
Several techniques such as this are implemented in the Python information theory package \texttt{dit}~\cite{dit}.

\section{Examples \& Comparisons}
\label{sec:examples}

We now demonstrate the behavior of our measure \Idep[]{} on a variety of source-target examples.
In simple cases---\textsc{Rdn}, \textsc{Syn}, \textsc{Copy}~\cite{griffith2014quantifying}---\Idep[]{} agrees with \Iproj[]{}, \Ibroja[]{}, and \Iccs[]{}.
There are, however, distributions where \Idep[]{} differs from the rest.
We concentrate on those.

\begin{figure}
  \centering
  \includegraphics{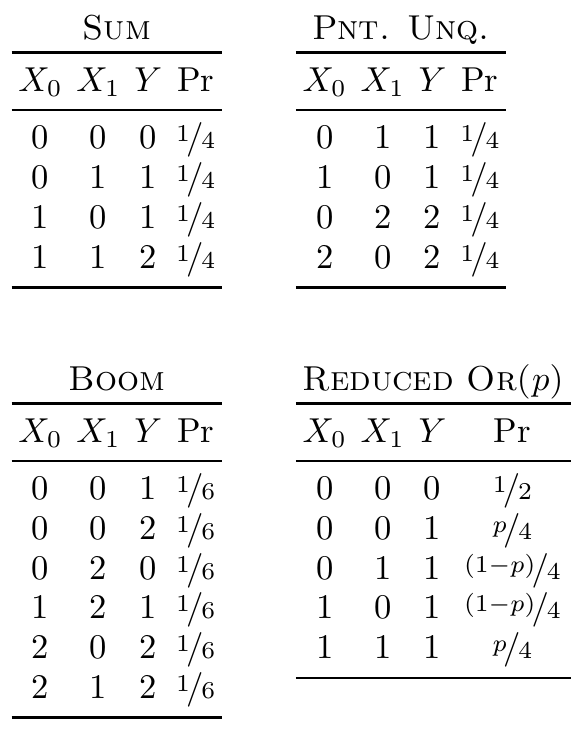}
  \caption{
    Four distributions of interest:
    \textsc{Sum} is constructed with \source{0}, \source{1} as independent binary variables while \target is their sum.
    \textsc{Pnt. Unq.} is from Ref.~\cite{finn2018pointwise}.
    \textsc{Boom} was found through a numeric search for distributions satisfying certain properties; namely, that \Iproj[]{} and \Ibroja[]{} differ.
    \textsc{Reduced Or} is adapted from Ref.~\cite{ince2017measuring}.
  }
  \label{tab:dists}
\end{figure}

\begin{table}
  \begin{tabular}[t]{ccccccc}
    \toprule
    & $\partial$ & \Imin[]{}         & \Iproj[]{}        & \Ibroja[]{}       & \Iccs[]{}         & \Idep[]{} \\
    \midrule
    \parbox[t]{2mm}{\multirow{4}{*}{\rotatebox[origin=c]{90}{\textsc{Sum}}}}
    & $01$       & 1                 & 1                 & 1                 & $\nicefrac{1}{2}$ & 0.68872 \\
    & $0$        & 0                 & 0                 & 0                 & $\nicefrac{1}{2}$ & 0.31128 \\
    & $1$        & 0                 & 0                 & 0                 & $\nicefrac{1}{2}$ & 0.31128 \\
    & $0\sep1$   & $\nicefrac{1}{2}$ & $\nicefrac{1}{2}$ & $\nicefrac{1}{2}$ & 0                 & 0.18872 \\
    \midrule
    \parbox[t]{2mm}{\multirow{4}{*}{\rotatebox[origin=c]{90}{\footnotesize\textsc{Pnt. Unq.}}}}
    & $01$       & $\nicefrac{1}{2}$ & $\nicefrac{1}{2}$ & $\nicefrac{1}{2}$ & 0                 & $\nicefrac{1}{4}$ \\
    & $0$        & 0                 & 0                 & 0                 & $\nicefrac{1}{2}$ & $\nicefrac{1}{4}$ \\
    & $1$        & 0                 & 0                 & 0                 & $\nicefrac{1}{2}$ & $\nicefrac{1}{4}$ \\
    & $0\sep1$   & $\nicefrac{1}{2}$ & $\nicefrac{1}{2}$ & $\nicefrac{1}{2}$ & 0                 & $\nicefrac{1}{4}$ \\
    \midrule
    \parbox[t]{2mm}{\multirow{4}{*}{\rotatebox[origin=c]{90}{\textsc{Boom}}}}
    & $01$       & 0.29248           & 0.29248           & 0.12581           & 0.12581           & 0.08781 \\
    & $0$        & $\nicefrac{1}{6}$ & $\nicefrac{1}{6}$ & $\nicefrac{1}{3}$ & $\nicefrac{1}{3}$ & 0.37133 \\
    & $1$        & $\nicefrac{1}{6}$ & $\nicefrac{1}{6}$ & $\nicefrac{1}{3}$ & $\nicefrac{1}{3}$ & 0.37133 \\
    & $0\sep1$   & $\nicefrac{1}{2}$ & $\nicefrac{1}{2}$ & $\nicefrac{1}{3}$ & $\nicefrac{1}{3}$ & 0.29533 \\
    \bottomrule
  \end{tabular}
  \caption{
    Partial information decomposition of the \textsc{Sum}, \textsc{Pnt. Unq.}, and \textsc{Boom} distributions.
  }
\label{tab:pids}
\end{table}

Consider the \textsc{Reduced Or($0$)} and \textsc{Sum} distributions~\cite{ince2017measuring} in Fig.\nobreakspace \ref {tab:dists}.
For these \Imin[]{}, \Iproj[]{}, and \Ibroja[]{} all compute no unique information.
Reference~\cite{ince2017measuring} provides an argument based on game theory that the channels $\source{0} \Rightarrow \target$ and $\source{1} \Rightarrow \target$ being identical (a special case of the Blackwell property \textbf{(BP)}~\cite{rauh2017extractable}) does not imply that unique information must vanish.
Specifically, the argument goes, the optimization performed in computing
\Ibroja[]{} \emph{artificially correlates} the sources, though this
interpretation is dependent upon the perspective one takes when considering the
PID~\cite{james2018perspective}.
One can interpret this as a sign that redundancy is being overestimated.
In these instances, \Idep[]{} qualitatively agrees with \Iccs[]{}, though they differ somewhat quantitatively.
See Table\nobreakspace \ref {tab:pids} for the exact values.

Reference~\cite{bertschinger2014quantifying} proves that \Iproj[]{} and \Ibroja[]{} are distinct measures.
The only example produced, though, is the somewhat opaque \textsc{Summed Dice} distribution.
Here, we offer \textsc{Boom} found in Fig.\nobreakspace \ref {tab:dists} as a more concrete
example to draw out such differences.\footnote{Although, it is not hard to find
a simpler example. See the \texttt{dit}~\cite{dit} documentation for another: \url{http://docs.dit.io/en/latest/measures/pid.html\#and-are-distinct}.}
Table\nobreakspace \ref {tab:pids} gives the measures' decomposition values.
Interestingly, \Imin[]{} agrees with \Iproj[]{}, while \Iccs[]{} agrees with \Ibroja[]{}.
\Idep[]{}, however, is distinct. All measures assign nonzero values to all four partial informations.
Thus, it is not clear if any particular method is superior in this case.

\begin{figure}
  \centering
  \includegraphics{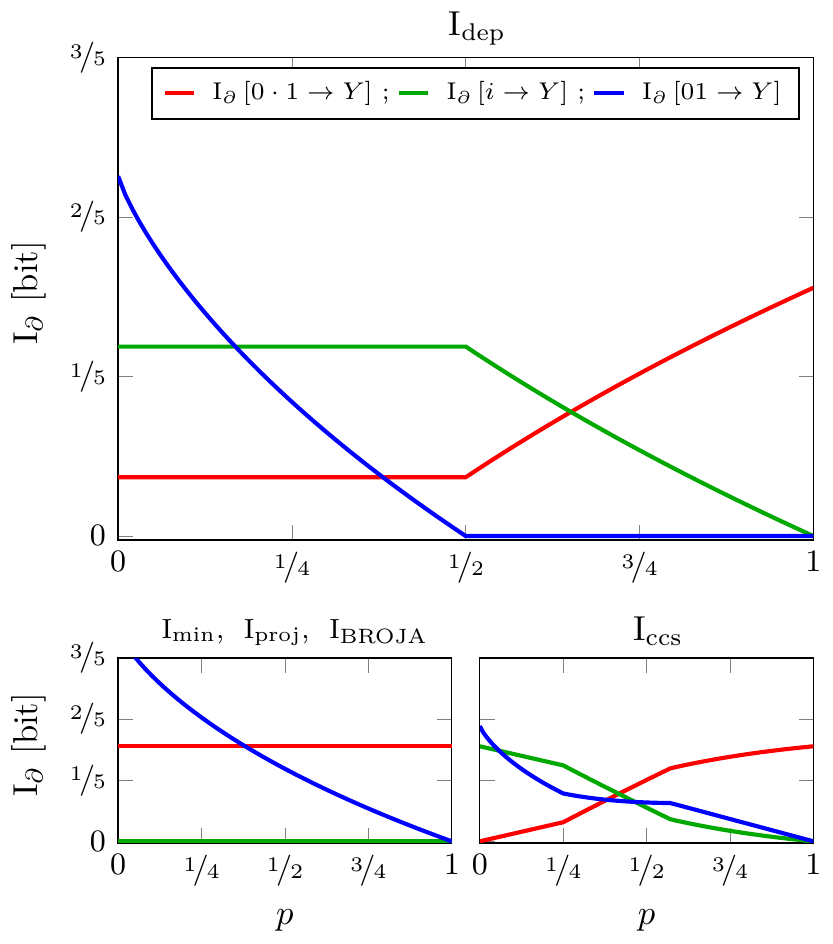}
\caption{
  Partial information decomposition of \textsc{Reduced Or($p$)} as a function of $p$: The \Idep[]{} decomposition shows an abrupt change in character at $p=\nicefrac{1}{2}$, corresponding to independent source-target channels switching from \emph{underestimating} the target distribution to \emph{overestimating}.
  Under \Ibroja[]{} (and \Imin[]{} and \Iproj[]{}), the redundant and unique components do not vary since the source-target marginals are invariant with $p$.
  The \Iccs[]{} decomposition exhibits two $p$ values of nonsmoothness, each corresponds to a change in sign of a coinformation component.
	}
\label{fig:reducedor_plots}
\end{figure}

Finally, consider the parametrized \textsc{Reduced Or($p$)} distribution, given in Fig.\nobreakspace \ref {tab:dists}.
Figure\nobreakspace \ref {fig:reducedor_plots} graphs this distribution's decomposition for all measures.
\Imin[]{}, \Iproj[]{}, and \Ibroja[]{} all produce the same decomposition as a function of $p$.
\Iccs[]{} and \Idep[]{} differ from those three and each other.
\Imin[]{}'s, \Iproj[]{}'s, and \Ibroja[]{}'s evaluation of redundant and unique information is invariant with $p$.
And, in the cases of \Iproj[]{} and \Ibroja[]{} this is due to the source-target marginals being invariant with respect to $p$~\cite{bertschinger2014quantifying}.
We next argue that \Idep[]{}'s decomposition---and specifically the ``kink'' observed---is both intuitive and reasonable.

Generically, the source-target channels ``overlap'' independent of $p$, so there is some invariant component to the redundancy.
Furthermore, consider the form of some conditional distributions when the sources are `0': the distribution $\Prob\left(\target\right) = \left\{ 0: \nicefrac{1}{2}, 1: \nicefrac{1}{2} \right\}$, while $\Prob\left(\target | \source{0} = 0\right) = \Prob\left(\target | \source{1} = 0\right) = \left\{ 0: \nicefrac{2}{3}, 1: \nicefrac{1}{3} \right\}$.
Finally, $\Prob\left(Y | \source{0}\source{1} = 00\right) = \left\{ 0: \nicefrac{2}{2+p}, 1: \nicefrac{p}{2+p} \right\}$.
The source-target channels are independent of $p$, but the joint sources-target channel depends upon it.

Consider the case of the two channels, $\source{0} \Rightarrow \target$ and $\source{1} \Rightarrow \target$, operating and independently influencing the value of \target.
Observation that $\source{0} = 0$ takes $\Prob\left(\target = 0\right) = \nicefrac{1}{2}$ to $\Prob\left(\target = 0\right) = \nicefrac{2}{3}$---a factor of $\nicefrac{4}{3}$ larger, and similarly for $\source{1} = 0$.
Together, one would then observe that $\Prob\left(\target^\prime | \source{0}\source{1} = 00\right) = \left\{ 0: \nicefrac{4}{5}, 1: \nicefrac{1}{5} \right\}$.
That is, each channel ``pushes'' $\Prob\left(\target = 0\right)$ $\nicefrac{4}{3}$ of the way from `0' toward `1'.
This independent ``pushing'' occurs exactly at $p = \nicefrac{1}{2}$.
For $p \geq \nicefrac{1}{2}$, this independence assumption \emph{overestimates} the probability of $\target = 0$.
That is, there is additional redundancy between the two channels.
For $p \leq \nicefrac{1}{2}$, the ``pushes'' from the two channels do not account for the true probability of $\target = 0$.
That is, synergistic effects occur.
\Idep[]{} cleanly reveals both of these features, while \Imin[]{}, \Iproj[]{}, and \Ibroja[]{} miss them and \Iccs[]{}'s multiple kinks makes it appear oversensitive.

\section{Discussion}
\label{sec:discussion}

We next describe several strengths of our \Idep[]{} measure when interpreting the behavior of the sources-target mapping channel.
The PID applied to a joint distribution naturally depends on selecting which variables are considered sources and which target.
In some cases---the \textsc{Rdn} and \textsc{Syn} distributions of Ref.~\cite{griffith2014quantifying}---the values of redundancy and synergy are independent of these choices and, in a sense, can be seen as a property of the joint distribution itself.
In other cases---the \textsc{And} distribution of Ref.~\cite{griffith2014quantifying}---the values of redundancy and synergy are not readily apparent in the joint distribution.
The concept of mechanistic redundancy---the existence of redundancy in spite of independent sources---is a manifestation of this.
What we term nonholistic synergy---synergy in the PID that does not arise from necessarily three-way interactions (that is, the third-order connected information~\cite{Schneidman2003}) in the distribution---is also due to the choice of sources and target.
We next discuss how the dependency decomposition and \Idep[]{} shed new insight into mechanistic redundancy and nonholistic synergy.

There are two aspects of the PID that do not directly reflect properties of the
joint distribution, but rather are determined by which variables are
selected as sources and which the target.
The first involves redundancy, where two sources may be independent but redundantly influence the target.
The second involves synergy, where there may be a lack of information at the triadic level of three-way interdependency, yet the sources collectively influence the target.
The dependency decomposition and \Idep[]{} make these phenomena explicit.

\subsection{Source versus Mechanistic Redundancy}
\label{subsec:redundancy}

\begin{figure}
  \includegraphics{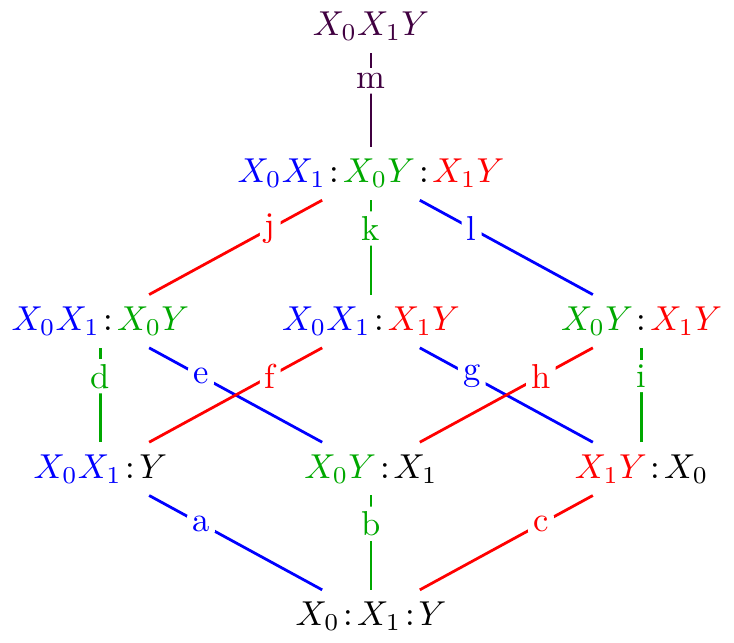}
  \caption{
    Dependency structure for two source variables \source{0} and \source{1} and one target variable \target.
    Edges colored blue correspond to adding constraint \source{0}\source{1}; edges colored green to adding constraint \source{0}\target; and edges colored red to \source{1}\target.
	  The unique information \Idep{\source{0}} is calculated by considering the least change in \I[\sigma]{\source{0}\source{1} : \target} along the green edges.
    See Appendix\nobreakspace \ref {app:dependency} and\nobreakspace Fig.\nobreakspace \ref {fig:reduceddependency} for identities among the edges important for \Idep[]{}.
	}
\label{fig:dependencystructure}
\end{figure}

An interesting concept within the PID domain is that of \emph{mechanistic redundancy}~\cite{harder2013bivariate}.
In its simplest form, this is existence of redundant information when the sources are independent.
The \textsc{And} distribution given in Table\nobreakspace \ref {tab:and} is a prototype for this phenomenon.
Though the two sources \source{0} and \source{1} are independent, all methods of quantifying partial information ascribe nonzero redundancy to this distribution.
Through the lens of \Idep[]{}, this occurs when the edge labeled l in Fig.\nobreakspace \ref {fig:dependencystructure} exceeds edge quantity $b - i = c - h$.
This means that the channels $\source{0} \Rightarrow \target$ and $\source{1} \Rightarrow \target$ are similar, so that when constraining just these two marginals the maximum entropy distribution artificially correlates the two sources.
This artificial correlation must then be broken when constraining the sources' marginal $\source{0}\source{1}$, leading to conditional dependence.
(Section\nobreakspace \ref {subsec:synergy} below draws out this implication.)

\begin{table}
  \anddist 
  \begin{tabular}[t]{cccccc}
    \toprule
    \textsc{And} & \Imin[]{} & \Iproj[]{} & \Ibroja[]{} & \Iccs[]{} & \Idep[]{} \\
    \midrule
    $01$     & $\nicefrac{1}{2}$ & $\nicefrac{1}{2}$ & $\nicefrac{1}{2}$ & 0.29248 & 0.27043 \\
    $0$      & 0                 & 0                 & 0                 & 0.20752 & 0.22957 \\
    $1$      & 0                 & 0                 & 0                 & 0.20752 & 0.22957 \\
    $0\sep1$ & 0.31128           & 0.31128           & 0.31128           & 0.10376 & 0.08170 \\
    \bottomrule
  \end{tabular}
  \caption{
    \textsc{And} distribution exemplifies both mechanistic redundancy and nonholistic synergy.
	}
\label{tab:and}
\end{table}

Mechanistic redundancy is closely tied to the concept of \emph{target monotonicity}~\cite{rauh2017extractable}:
\begin{itemize}
\item[\textbf{(TM)}]
  $\Icap{\source{0}\sep\source{1}} \ge
  \Icap[f(\target)]{\source{0}\sep\source{1}}$ ~.\\
	\propname{target monotonicity}
\end{itemize}
Said colloquially, taking a function of the target cannot increase redundancy.
However, one of the following three properties of a partial information measure must be false:
\begin{enumerate}
  \item \label{item:one} $\Icap[(\source{0}\source{1})]{\source{0}\sep\source{1}} = 0$,
  \item The possibility of mechanistic redundancy, or
  \item \label{item:three} Target monotonicity.
\end{enumerate}
In effect, any given method of quantifying the PID
cannot simultaneously assign zero redundancy to the ``two-bit copy'' distribution, allow mechanistic redundancy, and obey target monotonicity.
\Idep[]{} does not satisfy \textbf{(TM)}.
Reference~\cite{rauh2017extractable} demonstrated a general construction that
maps a redundancy measure not satisfying \textbf{(TM)} to one that does, in the
process violating property Item\nobreakspace \ref {item:one} above.

\subsection{Holistic versus Nonholistic Synergy}
\label{subsec:synergy}

A notion somewhat complementary to mechanistic redundancy is nonholistic synergy.
Holistic synergy, or the third-order connected information~\cite{Schneidman2003}, is the difference $\H{\source{0}\source{1}\target} - \H[\source{0}\source{1}:\source{0}\target:\source{1}\target]{\source{0}\source{1}\target}$.
This quantifies the amount of structure in the distribution that is only constrained by the full triadic probability distribution, not any subsets of marginals.
This quantity appears as the edge labeled m in Fig.\nobreakspace \ref {fig:dependencystructure}.
Nonholistic synergy, on the other hand, is synergy that exists purely from the bivariate relationships within the distribution.
This appears as $k - \min\{b, i, k\} = j - \min\{c, h, j\}$ in Fig.\nobreakspace \ref {fig:dependencystructure}.
This quantity has a natural interpretation: how much does the constraint $\source{0}\target$ influence \I{\source{0}\source{1} : \target} in the context of the other dyadic relationships ($\source{0}\source{1}$, $\source{1}\target$), minus the unique information \Idep{0}.
The total PID synergy is then $\Idep{01} = m + k - \min\{b, i, k\} = m + j - \min\{c, h, j\}$.

Here, again, the \textsc{And} distribution seen in Table\nobreakspace \ref {tab:and} exemplifies the phenomenon.
The \textsc{And} distribution is completely defined by the constraint $\source{0}\source{1}:\source{0}\target:\source{1}\target$.
That is, the \textsc{And} distribution is the only distribution satisfying these pairwise constraints.
This implies that the holistic synergy is zero.
In spite of this, all methods of quantifying partial information (correctly) assign nonzero synergy to this distribution.
This is a consequence of coinformation being negative.
This raises an interesting question: are there triadic (three-way) dependencies in the \textsc{And} distribution?
Notably, the distribution can be defined as the maximum entropy distribution satisfying certain pairwise marginals, yet it has negative co-information and therefore nonzero synergy and exhibits conditional dependence.

\subsection{Shortcomings}
\label{subsec:shortcomings}

\Idep[]{} comes with its own concerns, however.
First, it is defined using a minimum.
Besides being mildly aesthetically displeasing, this can lead to nondifferentiable (though continuous) behavior, as seen in Fig.\nobreakspace \ref {fig:reducedor_plots}.
Nondifferentiability can be seen as natural, as we have argued, if it coincides with a switch in qualitative behavior.

Perhaps more interestingly, \Idep[]{} does not correspond to either the camel
or elephant intuitions as described in Ref.~\cite{james2018perspective} which
proposes information-theoretical cryptography as a basis for unique information.
In the relatively straightforward example of the \textsc{Pnt. Unq.}
distribution, \Idep[]{} does not correspond with any other proposed method of
quantifying the PID.
In this instance, it is simple to state why \Idep[]{} quantifies the unique information as \SI{1/4}{\bit}: after constraining either $\source{0}\target$ or $\source{1}\target$, constraining the other only increases \I{\source{0}\source{1} : \target} by \SI{1/4}{\bit}; that is, the unique value of either $\source{0}\target$ or $\source{1}\target$ is only a quarter because there exist contexts where that is all it can contribute to \I{\source{0}\source{1} : \target}.
However, it is difficult to see the operational meaning of this value.
All other proposed methods match one or another secret key agreement rate.
And so, they at least  have a concrete operational interpretation.

Finally, \Idep[]{} is a measure of unique information and so it cannot alone be
used to quantify the PID with three or more sources.
And, in the event where unique informations in concert with standard
information measures are in fact sufficient for quantifying the entire
PID, \Idep[]{}'s adherence to the identity axiom implies that it necessarily does not obey local positivity with three or more sources.

\section{Conclusion}
\label{sec:conclusion}

We developed a promising new method \Idep[]{} of quantifying the partial information decomposition that circumvents many problems plaguing previous attempts.
It satisfies axioms \textbf{(S)}, \textbf{(SR)}, \textbf{(M)}, \textbf{(LP)}, and \textbf{(Id)}; see Appendix\nobreakspace \ref {app:properties}.
It does not, however, satisfy the Blackwell property \textbf{(BP)} and so, like \Iccs[]{}, it agrees with previous game-theoretic arguments raised in Ref.~\cite{ince2017measuring}.
Unlike \Iccs[]{}, though, \Idep[]{} satisfies \textbf{(LP)}.
This makes it the only measure satisfying \textbf{(Id)} and \textbf{(LP)} which does not require that redundancy is fixed by $\source{0}\target:\source{1}\target$.

The \Idep[]{} method does not overcome the negativity arising in the trivariate source explored in Refs.~\cite{bertschinger2013shared,rauh2014reconsidering,rauh2017secret}.
We agree with Ref.~\cite{rauh2017secret} that the likely solution is to
employ a different lattice.
We further believe that the flexibility of our dependency structure could lead
to methods of quantifying this hypothetical new lattice and to elucidating many
other challenges in decomposing joint information, especially once the
statistical significance of its structure applied to empirical data is
explored.

\section*{Acknowledgments}
\label{sec:acknowledgments}

The authors thank Robin Ince and Daniel Feldspar for many helpful conversations.
JPC thanks the Santa Fe Institute for its hospitality during visits.
JPC is an SFI External Faculty member.
This material is based upon work supported by, or in part by, the UC Davis Intel Parallel Computing Center, John Templeton Foundation grant 52095, Foundational Questions Institute grant FQXi-RFP-1609, the U.S. Army Research Laboratory and the U. S. Army Research Office under contracts W911NF-13-1-0390 and W911NF-13-1-0340.

\cleardoublepage

\appendix

\section{Constrained Three-Variable Maximum Entropy Distributions}
\label{app:constrained}

Here, we characterize the maximum entropy distributions defined in Eq.\nobreakspace \textup {(\ref {eq:argmaxent})} and used to populate the dependency structure.
In particular, we give maximum entropy distributions for forms of marginal constraints that describe the lowest three levels of the constraint lattice as shown in Fig.\nobreakspace \ref {fig:constraintlattice}.

Let us fix notation.
Consider a joint distribution $p(ABC)$ and maximum entropy distributions for some constraint set $\sigma$: $p_{\sigma}\left(ABC\right)$.
We label information measures and other quantities that are computed relative to this constrained maximum entropy distribution with the subscript $\sigma$: for example, $\H[A:B:C]{ABC}$ refers to the entropy of the product distribution $p(abc) = p(a)p(b)p(c)$, as this is the distribution consistent with the constraint $A:B:C$ and has maximum entropy.
Such quantities without a subscript are calculated from the original distribution.
In our use of the dependency structure to quantify unique information, we are interested in $\I[\sigma]{AB\!:\!C}$ as this will represent the sources-target mutual information.

\begin{figure}
  \subfloat[\label{subfig:a}]{
    \includegraphics{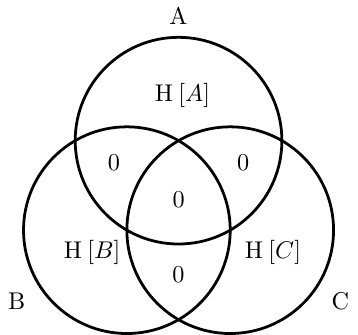}
  }
  \\
  \subfloat[\label{subfig:b}]{
    \includegraphics{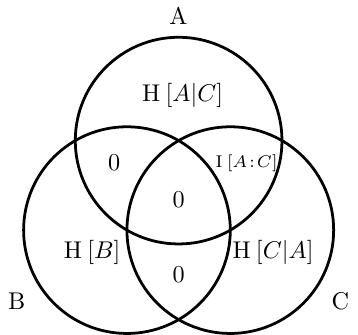}
  }
  \subfloat[\label{subfig:c}]{
    \includegraphics{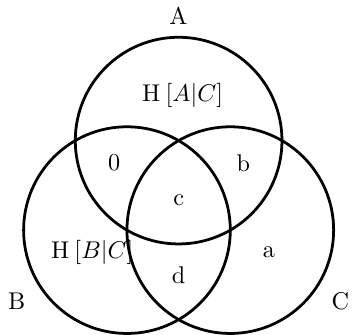}
  }
  \caption{
    Information diagrams corresponding to the maximum entropy distributions described in Eqs.\nobreakspace  \textup {(\ref {eq:nopairs})} to\nobreakspace  \textup {(\ref {eq:twopairs})} .
    The four variables in subfigure (c) satisfy $a+b+c+d=\H{C}$, $b+c=\I{A\!:\!C}$, and $c+d=\I{B\!:\!C}$.
	}
\label{fig:idiagrams}
\end{figure}

The lowest node in the dependency lattice constrains all single-variable marginal distributions, but no pairwise marginal distributions.
With only single-variable marginal distributions constrained, the maximum entropy distribution is such that the variables are independent, also known as the product distribution:
\begin{align}
  \label{eq:nopairs}
   p_{A:B:C}(ABC) = p(A)p(B)p(C)
   ~.
\end{align}
It can be seen that this must be the maximum entropy distribution, since an
increase in any mutual information corresponds to an equal decrease in at least two conditional entropies, resulting in a lower total entropy.
The informational structure of this distribution can be seen in Fig.\nobreakspace \ref {subfig:a}.

The first row up in the constraint lattice captures those antichains that contain one pairwise constraint and one single-variable constraint.
The maximum entropy distribution corresponding to this constraint set is given by:
\begin{align}
  \label{eq:onepair}
  p_{AC:B}(ABC) = p(AC)p(B)
  ~.
\end{align}
All atoms of the information diagram that capture the overlap of $\H{AC}$ and $\H{B}$ vanish.
Again, it can be seen that the maximum entropy distribution must take this form, since any deviation from the information partitioning seen in Fig.\nobreakspace \ref {subfig:b}, which satisfies the constraint $AC:B$, must result in an overall decrease to the entropy.

The elements in the second row of the constraint lattice include constraints on two pairwise marginal distributions each.
These constraints both contain one of the variables, and the maximum entropy distribution takes on the following Markov form:
\begin{align}
  \label{eq:twopairs}
  p_{AC:BC}(ABC) = p(A|C)p(B|C)p(C)
  ~.
\end{align}
This distribution forms a Markov chain $A \markov C \markov B$ and therefore $\I[AC:BC]{A\!:\!B|C}=0$.
To see that this must be the form of the maximum entropy distribution, consider the expansion:
\begin{align*}
  \H{ABC} = \H{C} + \H{A|C} + \H{B|C} - \I{A\!:\!B|C}
  ~.
\end{align*}
The first three terms of the righthand side are constrained by $p(C)$, $p(AC)$, and $p(BC)$, respectively.
Since the conditional mutual information is necessarily nonnegative, the final term being zero corresponds to the distribution with the maximum entropy.
Such a distribution is realized by the given Markov chain.
The mutual information $\I[AC:BC]{AB\!:\!C}$ is equal to:
\begin{align*}
  \I{A:C} + \sum_{\substack{a \in \mathcal{A} \\ b \in \mathcal{B} \\ c \in \mathcal{C}}} p(a, b, c) \log_2{\frac{p(a)p(b, c)}{p(c)p(a, b)}}
  ~.
\end{align*}

The information diagrams for each of these three distributions are given in Fig.\nobreakspace \ref {fig:idiagrams}.

With the structure of these distributions in hand, we now turn to proving several properties of the \Idep[]{} measure.

\section{Sources-Target Dependency Structure}
\label{app:dependency}

\begin{figure}
  \includegraphics{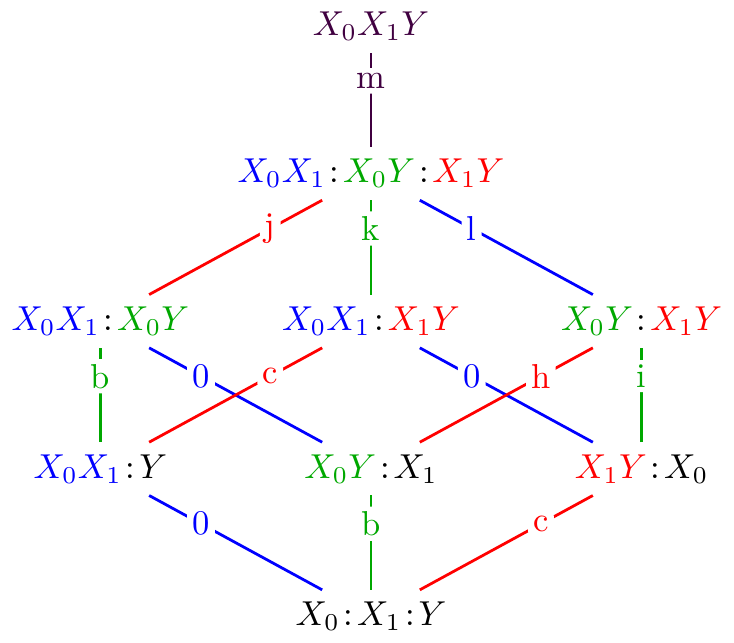}
  \caption{
    The reduced form of the dependency lattice quantified by sources-target mutual information, with $b=\I{\source{0}\!:\!\target}$ and $c=\I{\source{1}\!:\!\target}$.
    All edges are guaranteed to be nonnegative except for $l$.
  }
  \label{fig:reduceddependency}
\end{figure}

Interpreting the set of antichain covers as possible marginal constraints on probability distributions, we defined a dependency lattice that gradually introduces dependencies into an otherwise unstructured distribution.
In this method of quantifying the partial information decomposition, \Idep[]{} is defined according to the node-node differences of the sources-target mutual information $\I{\source{0}\source{1} : \target}$.
These lattice edges are labeled in Fig.\nobreakspace \ref {fig:dependencystructure}.

The maximum entropy distributions on the lowest three levels of nodes follow the forms given in Appendix\nobreakspace \ref {app:constrained}.
By appropriately assigning $\source{0}$, $\source{1}$, and $\target$ to $A$,
$B$, and $C$, we obtain the relationships arising among the edges summarized in Fig.\nobreakspace \ref {fig:reduceddependency} and the following results for the sources-target mutual information:
\begin{align}
\label{eq:s-tI1}\I[\source{0}:\source{1}:\target]{\source{0}\source{1}\!:\!\target}
	&=0 \\
\label{eq:s-tI2}\I[\source{0}\source{1}:\target]{\source{0}\source{1}\!:\!\target}
	&=0 \\
\label{eq:s-tI3}\I[\source{0}\target:\source{1}]{\source{0}\source{1}\!:\!\target}
	&=\I{\source{0}\!:\!\target} \\
\label{eq:s-tI4}\I[\source{1}\target:\source{0}]{\source{0}\source{1}\!:\!\target}
	&=\I{\source{1}\!:\!\target} \\
\label{eq:s-tI5}\I[\source{0}\source{1}:\source{0}\target]{\source{0}\source{1}\!:\!\target}
	&=\I{\source{0}\!:\!\target}  \\
\label{eq:s-tI6}\I[\source{0}\target:\source{1}\target]{\source{0}\source{1}\!:\!\target}
	&=\I{\source{1}\!:\!\target}
\end{align}
Now, we can simply read off the values of $a$ through $g$, and derive other simple relationships:
\begin{align*}
  a & =e=g=0\\
  d & =b=\I{\source{0}\!:\!\target}\\
  f & =c=\I{\source{1}\!:\!\target}\\
  b+h & =c+i\\
  b+j+m  & = c+k+m = \I{\source{0}\source{1}\!:\!\target}
  ~.
\end{align*}

Furthermore:
\begin{align*}
  h &= \I[\source{0}\target:\source{1}\target]{\source{0}\source{1} : \target} - \I{\source{0} : \target} \\
    &= \I[\source{0}\target:\source{1}\target]{\source{1} : \target | \source{0}} \\
  i &= \I[\source{0}\target:\source{1}\target]{\source{0}\source{1} : \target} - \I{\source{1} : \target} \\
    &= \I[\source{1}\target:\source{1}\target]{\source{0} : \target | \source{1}} \\
  j &= \I[\source{0}\source{1}:\source{0}\target:\source{1}\target]{\source{0}\source{1} : \target} - \I{\source{0} : \target} \\
    &= \I[\source{0}\source{1}:\source{0}\target:\source{1}\target]{\source{1} : \target | \source{0}} \\
  k &= \I[\source{0}\source{1}:\source{0}\target:\source{1}\target]{\source{0}\source{1} : \target} - \I{\source{1} : \target} \\
    &= \I[\source{0}\source{1}:\source{1}\target:\source{1}\target]{\source{0} : \target | \source{1}}
  ~.
\end{align*}

Our first task is to demonstrate that all edges, save the one labeled $l$, correspond to nonnegative differences in the sources-target mutual information.
The edges $b$ and $c$ are given by mutual informations and so must be nonnegative.
The edges $h$, $i$, $j$, and $k$ are given by conditional mutual informations
computed relative to certain maximum entropy distributions and, therefore, must also be nonnegative.
The edge labeled $m$ is the third-order connected
information~\cite{Schneidman2003}, which can be written as a Kullback-Leibler divergence and so must also be nonnegative.
This leaves only the edge $l$ potentially negative.
These last two edges, $l$ and $m$, do not involve the addition of a source-target constraint and so are not considered in computing \Idep[]{}.

Next, we demonstrate that only two edges meaningfully contribute to the determination of \Idep[]{}.
Since the coinformation $\I[\source{0}\target:\source{1}\target]{\source{0}:\source{1}:\target} = \I[\source{0}\target:\source{1}\target]{\source{0} : \source{1}}$ is necessarily nonnegative, we find that:
\begin{align*}
  h &\leq \I{\source{1} : \target} = c\\
  i &\leq \I{\source{0} : \target} = b
  ~.
\end{align*}
And so, in computing \Idep[]{} one need only consider the edges i and k (for \Idep{\source{0}}) or h and j (for \Idep{\source{1}}).

\section{Bivariate Partial Information \Idep[]{} Decomposition}
\label{app:properties}

This section establishes the properties of the bivariate partial information decomposition induced by \Idep[]{}.

\subsection*{Self-redundancy}
\label{app:selfredundancy}

Property \textbf{(SR)}: $\Icap{i} = \I{\source{i}\!:\!\target}$.

The shared information for an antichain with a single subset of source variables is precisely the mutual information between those source variables and the target.

We take this axiom constructively, defining three of the four shared informations \Icap[]{} and providing three constraints on partial informations \Ipart[]{} in Eqs.\nobreakspace  \textup {(\ref {eq:bivariatea})} to\nobreakspace  \textup {(\ref {eq:bivariatec})} .

\subsection*{Nonnegativity}
\label{app:nonnegativity}

Property \textbf{(LP)}: For all antichains $\sigma$:
$\Ipart{\sigma}\ge 0$.

Every partial information value resulting from the mobius inversion of the redundancy lattice is nonnegative.

We begin with the unique partial information from \source{0}.
Both arguments of the minimum in \Idep{0} were shown to be nonnegative in Appendix\nobreakspace \ref {app:dependency}:
\begin{align*}
  \Ipart{0} = \Idep{0} = \min\left(i,k\right) \ge 0~.
\end{align*}

Using the self-redundancy axiom to define $\Icap{0} = \I{\source{0}\!:\!\target}$ and knowing that $\Ipart{0} = \min\left(i,k\right)\le\I{\source{0}\!:\!\target}$ from Appendix\nobreakspace \ref {app:dependency}, we have:
\begin{align*}
  \Ipart{0\sep1}  = \Icap{0} - \Ipart{0} \ge 0~.
\end{align*}

To determine the signs of the remaining two partial information atoms, we must consider the ordering of $i$ and $k$.

Reductions are done by using results of Appendix\nobreakspace \ref {app:dependency}.
We repeatedly use the redundancy lattice inversions: $\Ipart{1} = \Icap{1} - \Ipart{0\sep1}$ and $\Ipart{01} = \Icap{01} - \Ipart{0\sep1} - \Ipart{0} - \Ipart{1}$.

CASE 1: $i \le k$.
\begin{align*}
  \Ipart{1} &= c - \left( b - i \right) = h \\
            & \ge 0 ~,\\
  \Ipart{01} &= \left( c + k + m \right) - \left( c - h \right) - i - h \\
             &= m + k - i \\
             & \ge m \\
			 & \ge 0
  ~.
\end{align*}

CASE 2: $k \le i$.
\begin{align*}
  \Ipart{1} &= c - \left(b-k\right) \\
            &= j \\
            & \ge 0 ~,\\
  \Ipart{01} & = \left( b + j + m \right) - \left( b - k \right) - k - j \\
             & = m \\
			 & \ge 0
  ~.
\end{align*}
In each case, the second unique information is found to be equivalent to another nonnegative edge in the dependency lattice.
Additionally, the synergy is found to be bounded from below by the ``holistic'' synergy $m$.

\subsection*{Monotonicity}
\label{app:monotonicity}

Property \textbf{(M)}: $\alpha \preceq \beta \implies \Icap{\alpha} \leq \Icap{\beta}$.

For any two antichains $\alpha$, $\beta$, an ordering between them implies the same ordering of their shared informations.

This follows immediately from \textbf{(LP)} above.

\subsection*{Symmetry}
\label{app:symmetry}

Property \textbf{(S)}: Under source reorderings, the following is invariant:
\begin{align*}
  \Ipart{0}
  ~.
\end{align*}
The dependency lattice is symmetric by design.
Relabeling the random variables is equivalent to an isomorphic relabeling of the lattice.
Therefore, we consider the effect of completing the partial information decomposition by either \Idep{0} or \Idep{1}.

Computing $\Idep{0} = \min\left(b,i,k\right)$ gives $\Ipart{1}=\min\left(c,h,j\right)$, although we never explicitly do the second minimization.
This requires simple algebra from the various multiple-paths constraints given in Appendix\nobreakspace \ref {app:dependency}.
In each of the Appendix\nobreakspace \ref {app:symmetry} cases, \Ipart{1} was found to be one of $\{c,h,j\}$.
Straightforward algebra shows that it is necessarily the minimum of them in each of the particular cases.

\subsection*{Identity}
\label{app:identity}

Property \textbf{(Id)}:
\begin{align*}
  \Ipart[\source{0}\source{1}]{0\sep1} = \I{\source{0}\!:\!\source{1}}
  ~.
\end{align*}
Consider sources $\source{0}$ and $\source{1}$ and output $\target = \source{0}\source{1}$, the concatenation of inputs.
The mutual information of either source with the target is simply the entropy of that source.
That is, $b=\H{\source{0}}$.
Using appropriate permutations of Eqs.\nobreakspace \textup {(\ref {eq:onepair})} and\nobreakspace  \textup {(\ref {eq:twopairs})} ($A=\source{0},\,B=\target,\,C=\source{1}$), we find that $i=\H{X_0|X_1}$.
Now, starting at the constraint $\sigma=\source{0}\source{1}:\source{1}\target$ as in Eq.\nobreakspace \textup {(\ref {eq:twopairs})} ($A=\source{0},\,B=\source{1},\,C=\target$), we see that additionally constraining $p(\source{0}\target)$ fully constrains the distribution to its original form, with a sources-target mutual information of $\H{\source{0}\source{1}}$.
That is, $k=\H{\source{0}|\source{1}}$.
The minimum of these three quantities gives $\Idep{0} = \H{\source{0}|\source{1}}$ and therefore verifies the identity axiom.


\begin{thebibliography}{10}

\bibitem{gates2016control}
A.~J. Gates and L.~M. Rocha.
\newblock Control of complex networks requires both structure and dynamics.
\newblock {\em Scientific reports}, 6:24456, 2016.

\bibitem{faber2018computation}
S.~P. Faber, N.~M. Timme, J.~M. Beggs, and E.~L. Newman.
\newblock Computation is concentrated in rich clubs of local cortical neurons.
\newblock {\em bioRxiv}, page 290981, 2018.

\bibitem{vijayaraghavan2017anatomy}
V.~Vijayaraghavan, R.~G. James, and J.~P. Crutchfield.
\newblock Anatomy of a spin: the information-theoretic structure of classical
  spin systems.
\newblock {\em Entropy}, 19(5):214, 2017.

\bibitem{james2018modes}
R.~G. James, B.~D.~M. Ayala, B.~Zakirov, and J.~P. Crutchfield.
\newblock Modes of information flow.
\newblock {\em arXiv preprint arXiv:1808.06723}, 2018.

\bibitem{arellano2013shannon}
R.~B. Arellano-Valle, J.~E. Contreras-Reyes, and M.~G. Genton.
\newblock Shannon entropy and mutual information for multivariate
  skew-elliptical distributions.
\newblock {\em Scandinavian Journal of Statistics}, 40(1):42--62, 2013.

\bibitem{Shan48a}
C.~E. Shannon.
\newblock A mathematical theory of communication.
\newblock {\em Bell Sys. Tech. J.}, 27:379--423, 623--656, 1948.

\bibitem{Shan56b}
C.~E. Shannon.
\newblock The bandwagon.
\newblock {\em IEEE Transactions on Information Theory}, 2(3):3, 1956.

\bibitem{Shan53a}
C.~E. Shannon.
\newblock The lattice theory of information.
\newblock {\em Trans. IRE Prof. Group Info. Th.}, 1(1):105--107, 1953.

\bibitem{Birk40a}
G.~Birkhoff.
\newblock {\em Lattice Theory}.
\newblock American Mathematical Society, Providence, Rhode Island, first
  edition, 1940.

\bibitem{Will10a}
P.~L. Williams and R.~D. Beer.
\newblock Nonnegative decomposition of multivariate information.
\newblock {\em arXiv:1004.2515}.

\bibitem{james2017multivariate}
R.~G. James and J.~P. Crutchfield.
\newblock Multivariate dependence beyond shannon information.
\newblock {\em Entropy}, 19(10):531, 2017.

\bibitem{harder2013bivariate}
M.~Harder, C.~Salge, and D.~Polani.
\newblock Bivariate measure of redundant information.
\newblock {\em Phys. Rev. E}, 87(1):012130, 2013.

\bibitem{griffith2014quantifying}
V.~Griffith and C.~Koch.
\newblock Quantifying synergistic mutual information.
\newblock In {\em Guided Self-Organization: Inception}, pages 159--190.
  Springer, 2014.

\bibitem{bertschinger2014quantifying}
N.~Bertschinger, J.~Rauh, E.~Olbrich, J.~Jost, and N.~Ay.
\newblock Quantifying unique information.
\newblock {\em Entropy}, 16(4):2161--2183, 2014.

\bibitem{chicharro2017quantifying}
D.~Chicharro.
\newblock Quantifying multivariate redundancy with maximum entropy
  decompositions of mutual information.
\newblock {\em arXiv:1708.03845}.

\bibitem{griffith2014intersection}
V.~Griffith, E.~K.P. Chong, R.~G. James, C.~J. Ellison, and J.~P. Crutchfield.
\newblock Intersection information based on common randomness.
\newblock {\em Entropy}, 16(4):1985--2000, 2014.

\bibitem{ince2017measuring}
R.~A.A. Ince.
\newblock Measuring multivariate redundant information with pointwise common
  change in surprisal.
\newblock {\em Entropy}, 19(7):318, 2017.

\bibitem{bertschinger2013shared}
N.~Bertschinger, J.~Rauh, E.~Olbrich, and J.~Jost.
\newblock Shared information-new insights and problems in decomposing
  information in complex systems.
\newblock In {\em Proceedings of the European Conference on Complex Systems
  2012}, pages 251--269. Springer, 2013.

\bibitem{rauh2014reconsidering}
J.~Rauh, N.~Bertschinger, E.~Olbrich, and J.~Jost.
\newblock Reconsidering unique information: Towards a multivariate information
  decomposition.
\newblock In {\em Information Theory (ISIT), 2014 IEEE International Symposium
  on}, pages 2232--2236. IEEE, 2014.

\bibitem{chicharro2016redundancy}
D.~Chicharro and S.~Panzeri.
\newblock Redundancy and synergy in dual decompositions of mutual information
  gain and information loss.
\newblock {\em arXiv:1612.09522}.

\bibitem{banerjee2015synergy}
P.~K. Banerjee and V.~Griffith.
\newblock Synergy, redundancy and common information.
\newblock {\em arXiv:1509.03706}.

\bibitem{rauh2017secret}
J.~Rauh.
\newblock Secret sharing and shared information.
\newblock {\em Entropy}, 19(11):601, 2017.

\bibitem{rauh2017extractable}
J.~Rauh, P.~K. Banerjee, E.~Olbrich, J.~Jost, and N.~Bertschinger.
\newblock On extractable shared information.
\newblock {\em Entropy}, 19(7):328, 2017.

\bibitem{krippendorff2009ross}
K.~Krippendorff.
\newblock {Ross Ashby's} information theory: {A} bit of history, some solutions
  to problems, and what we face today.
\newblock {\em Intl. J. General Systems}, 38(2):189--212, 2009.

\bibitem{zwick2004overview}
M.~Zwick.
\newblock An overview of reconstructability analysis.
\newblock {\em Kybernetes}, 33(5/6):877--905, 2004.

\bibitem{james2018perspective}
R.~G. James, J.~Emenheiser, and J.~P. Crutchfield.
\newblock A perspective on unique information: Directionality, intuitions, and
  secret key agreement.
\newblock {\em arXiv:1808.08606}.

\bibitem{Cover2006}
T.~M. Cover and J.~A. Thomas.
\newblock {\em {Elements of Information Theory}}.
\newblock Wiley-Interscience, New York, second edition, 2006.

\bibitem{MacKay2003}
D.~MacKay.
\newblock {\em Information Theory, Inference, and Learning Algorithms}.
\newblock Cambridge University Press, Cambridge, United Kingdom, 2003.

\bibitem{Yeung2008}
R.~W. Yeung.
\newblock {\em Information theory and network coding}.
\newblock Springer, New York, 2008.

\bibitem{dijkstra1982numbering}
E.~W. Dijkstra.
\newblock Why numbering should start at zero.
\newblock 1982.

\bibitem{Bell2003}
A.~J. Bell.
\newblock The co-information lattice.
\newblock In S.~Makino S.~Amari, A.~Cichocki and N.~Murata, editors, {\em Proc.
  Fifth Intl. Workshop on Independent Component Analysis and Blind Signal
  Separation}, volume ICA 2003, pages 921--926, New York, 2003. Springer.

\bibitem{maurer1999unconditionally}
U.~M. Maurer and S.~Wolf.
\newblock Unconditionally secure key agreement and the intrinsic conditional
  information.
\newblock {\em IEEE Transactions on Information Theory}, 45(2):499--514, 1999.

\bibitem{Jayn83}
E.~T. Jaynes.
\newblock Where do we stand on maximum entropy?
\newblock In E.~T. Jaynes, editor, {\em Essays on Probability, Statistics, and
  Statistical Physics}, page 210. Reidel, London, 1983.

\bibitem{virgodecomposing}
N.~Virgo and D.~Polani.
\newblock Decomposing multivariate information (abstract).
\newblock 2017.
\newblock Accessed 26 April 2018,
  \url{https://www.mis.mpg.de/fileadmin/pdf/abstract_gso18_3302.pdf}.

\bibitem{amari2001information}
S.~Amari.
\newblock Information geometry on hierarchy of probability distributions.
\newblock {\em IEEE transactions on information theory}, 47(5):1701--1711,
  2001.

\bibitem{Schneidman2003}
E.~Schneidman, S.~Still, M.~J. Berry, and W.~Bialek.
\newblock Network information and connected correlations.
\newblock {\em Phys. Rev. Lett.}, 91(23):238701, 2003.

\bibitem{runge2015quantifying}
J.~Runge.
\newblock Quantifying information transfer and mediation along causal pathways
  in complex systems.
\newblock {\em Phys. Rev. E}, 92(6):062829, 2015.

\bibitem{sun2014causation}
J.~Sun and E.~M. Bollt.
\newblock Causation entropy identifies indirect influences, dominance of
  neighbors and anticipatory couplings.
\newblock {\em Physica D: Nonlinear Phenomena}, 267:49--57, 2014.

\bibitem{Pear88a}
J.~Pearl.
\newblock {\em Probabilistic Reasoning in Intelligent Systems}.
\newblock Morgan Kaufmann, New York, 1988.

\bibitem{dit}
R.~G. James, C.~J. Ellison, and J.~P. Crutchfield.
\newblock {dit}: a {P}ython package for discrete information theory.
\newblock {\em The Journal of Open Source Software}, 3(25):738, 2018.

\bibitem{finn2018pointwise}
C.~Finn and J.~T. Lizier.
\newblock Pointwise partial information decomposition using the specificity and
  ambiguity lattices.
\newblock {\em Entropy}, 20(4):297, 2018.

\end{thebibliography}
\end{document}